# GEOMScope: Large Field-of-view 3D Lensless Microscopy with Low Computational Complexity

Feng Tian[1], Junjie Hu[1], and Weijian Yang[1]*

[1]Department of Electrical and Computer Engineering, University of California, Davis, CA 95616, USA
*wejyang@ucdavis.edu

**ABSTRACT**
Recent development of lensless imagers has enabled three-dimensional (3D) imaging through a thin piece of optics in close proximity to a camera sensor. A general challenge of wide-field lensless imaging is the high computational complexity and slow speed to reconstruct 3D objects through iterative optimization process. Here, we demonstrated GEOMScope, a lensless 3D microscope that forms image through a single layer of microlens array and reconstructs objects through a geometrical-optics-based pixel back projection algorithm and background suppressions. Compared to others, our method allows local reconstruction, which significantly reduces the required computation resource and increases the reconstruction speed by orders of magnitude. This enables near real-time object reconstructions across a large volume of 23×23×5 mm$^3$, with a lateral resolution of 40 μm and axial resolution of 300 μm. Our system opens new avenues for broad biomedical applications such as endoscopy, which requires both miniaturized device footprint and real-time high resolution visualization.

**INTRODUCTION**
Recent development of lensless light field imaging devices have enabled compact three-dimensional (3D) imaging systems through single exposure. Lensless imagers replace the bulk lenses in conventional imaging systems by a thin amplitude or phase mask [1-8]. While fine resolutions across a large 3D volume have been reported, lensless imagers face challenges on their expensive computational cost in object reconstruction and tradeoffs among reconstruction quality, speed and volume. Lensless imagers reconstruct objects globally by solving the inverse problem of imaging through convex optimizations [1, 3-8] or deep neural networks [9-11]. Both require a large amount of computational resource that scales with both the image size and object size. This limitation stems from that the object space has to be reconstructed altogether. For microscopy applications where the sample is relatively sparse, such a global reconstruction results in a lot of redundant voxel-to-pixel mappings and memory-consuming system matrix operations. The high demand of computation not only slows the reconstruction speed but also limits the total number of object voxels that can be reconstructed [12-15].

Here we report a fluorescence microscope using a single layer of random microlens array and a hybrid 3D reconstruction algorithm, which can reduce the computation resource by orders of magnitude (Fig. 1). The hybrid reconstruction combines a pixel back projection algorithm and a background suppression algorithm. Compared with the commonly used iterative optimization techniques that reconstruct the object globally, our approach solves object information locally and independently through small patches of images. In our approach, the computation resources required for solving *single* object feature is fixed, and the overall computation cost scales linearly with the voxel number. This results in orders of magnitude less computational memory and orders of magnitude faster processing speed (Fig. 1E, Supplementary Table S1).

We term our microscope to be "GEOMScope" as the pixel back projection algorithm utilizes ray tracing and geometrical priors of microlens pattern for local object reconstruction. The pixel back projection algorithm first builds the point spread function (PSF) based on the microlens arrangement and the geometrical ray-projection (Fig. 1A-B). The raw captured image is then back projected to the object domain based on the PSF to retrieve the intensity distribution of the object (Fig. 1C). To suppress background light from unfocused objects, we use either a point clustering method or a convolutional neural network, to post process the initially reconstructed results (Fig. 1D). Such a reconstruction strategy is proven efficient and capable of real-time and high-resolution reconstruction.

We emphasize that our reconstruction through geometrical optics does not degrade the resolution, and we can achieve similar resolution as the iterative optimization techniques. In conventional wisdom, reconstruction through geometrical optics often results in a lower resolution than iterative optimization process, particularly in microscopic imaging [14, 16, 17]. This stems from that iterative optimizations often use a more precise forward model and PSF than geometrical optics. Here, we engineer GEOMScope such that the PSF is sparse and can be considered as a Dirac comb function. As a result, geometrical optics can leverage a precise image formation model, resulting in a resolution on par with iterative optimization. This is also the key difference between GEOMScope and the early light field microscopy using ray tracing approach [18] (Fig. 1F). In the latter case, reconstruction through ray tracing fails to capture the complex PSF, leading to a poor lateral resolution equivalent to the size of a single microlens unit [18, 19].

The focused plenoptic camera [20] improves this lateral resolution by tightening the PSF, similar as our strategy. It reconstructs the object by stitching the image patches from individual sub-image of each microlens. While this works well for macroscopic photography, it fails in microscopy application as the much larger magnification leads to substantial crosstalk between the sub-images. Our pixel-back projection reconstructs object features by collecting contributive image pixels based on local PSF, and can handle large sub-image crosstalk (Fig. 1G).

Another feature of our reconstruction algorithm is that it can well handle complex objects, thought it requires low computation resources. This is attributed to the background suppression step after the pixel back projection. Our geometrical optics approach can reconstruct objects at different depths, as if they were imaged sequentially by mechanically adjusting the distance between the object and the lens in a conventional imaging system (Fig. 1C). In other words, the reconstruction contains defocused light as the background. This is inevitable because of the local reconstruction nature of pixel back projection. This problem is resolved through the subsequent background suppression algorithms, which greatly sharpens the features (Fig. 1D). Such a hybrid approach enables imaging and faithful reconstructing of complex volumetric objects.

Thanks to the high efficiency and high resolution reconstruction, GEOMScope enables a large field-of-view imaging in high resolution, with a near real-time imaging reconstruction speed. Using a single layer of microlens array (20×20 mm$^2$), GEOMScope is capable of single-shot 3D imaging across a large volume of ~23×23×5 mm$^3$ with a lateral resolution of ~40 μm and an axial resolution of ~300 μm. A total of ~5.5×10$^6$ resolvable points can be reconstructed in 10s of second scale, which was not possible using existing approaches. We experimentally validated the 3D microscope through volumetric imaging of fluorescence beads and objects across a large volume. GEOMScope is particularly promising for biomedical applications that require real-time high resolution 3D visualization through miniaturized and implantable imaging devices.

In the following, we first present the imaging formation process and object reconstruction algorithm of GEOMScope. We then show the microlens array design, followed by the simulated results of system PSF, resolution, and 3D object reconstruction. Next, we verify the 3D resolution of the fabricated device by experimentally measuring point sources and imaging of USAF resolution targets. We demonstrate the 3D resolving ability with fluorescent particles and phosphorescent objects. Lastly, we show the computation speed improvement and reconstruction quality of GEOMScope compared to the prevailing reconstruction approaches at various scales of data sizes.

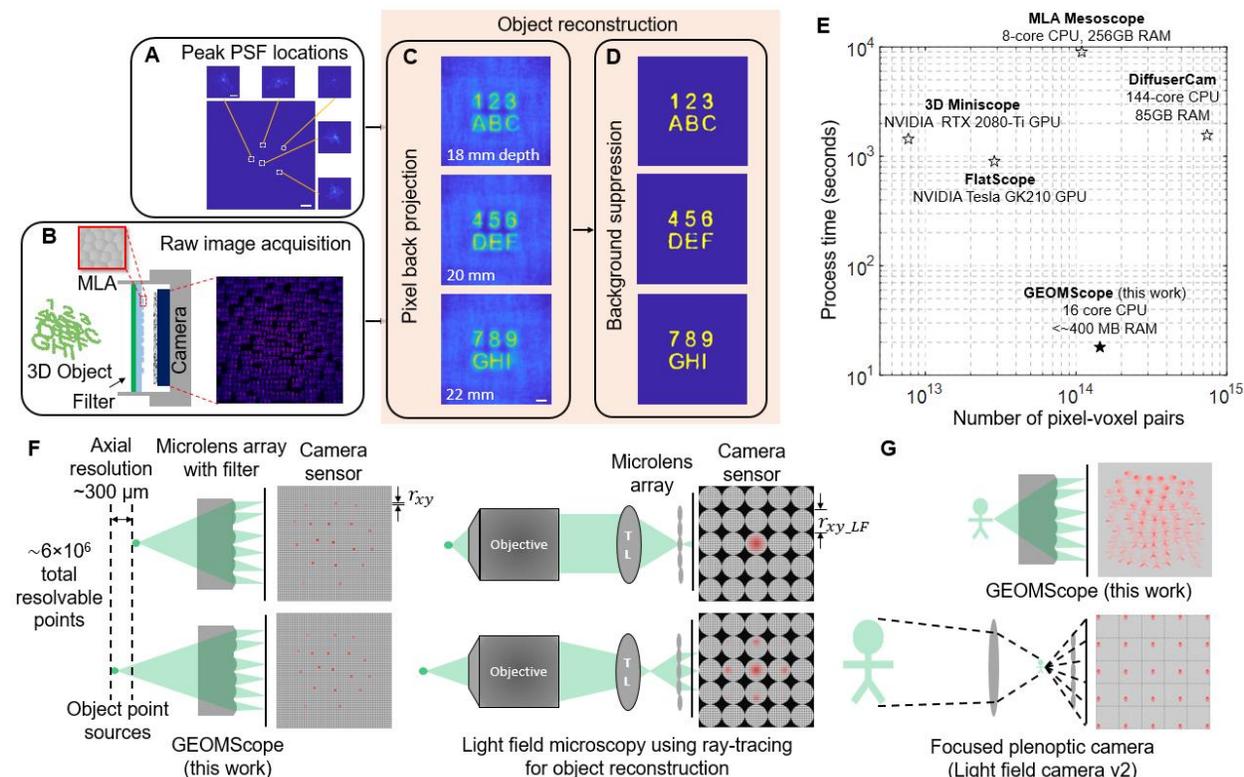

**Fig. 1. Architecture of GEOMScope.** (A) Experimental PSF measurement from single point source. Scale bar: 2 mm; inset, 200 μm. (B) Schematic of integrated 3D fluorescent imager. Raw image is simulated through ray tracing in OpticStudio. MLA, microlens array. (C) Initial reconstruction at different depths through pixel back projection. (D) Resolved objects with background suppression through a trained neural network. (E)

Comparison between GEOMScope and other lensless imagers (FlatScope [2], DiffuserCam [1], 3D Miniscope [5], MLA Mesoscope [7]) on the number of camera pixel and object voxel pair versus computation time (with the process type and required RAM) in reconstruction. The number of object voxel is either the reported value or the derived value as the ratio of imaged volume and 3D resolution. (F) Comparison between GEOMScope (left) and traditional light-field microscope using ray-tracing for object reconstruction (right). The resolution in traditional light-field microscope is limited by the size of microlenses. TL, tube lens. (G) Comparison between GEOMScope (top) and focused plenoptic camera (i.e. light field camera v2, bottom). Existing geometric optics based reconstruction algorithm in focused plenoptic camera only works when the magnification is small (i.e. macroscopic imaging in photography setting) and there is no substantial overlap between sub-images from each microlens unit.

## RESULTS
### Image formation

The system architecture of GEOMScope contains the microlens array with a filling factor of ~1, fluorescence emission filter, and the image sensor (Fig. 2A). The microlens array and the image sensor has a similar lateral size (on the orders of 10s mm) and are located in close proximity (with a distance of $v$ on the orders of a few mm), so the overall geometry of the integrated imager is flat. Each lens unit is randomly positioned in the microlens array (Fig. 1B). The PSF is thus sparse and non-periodic (Fig. 1A), which reduces the background light and lens unit crosstalk. Although the PSF of the entire microlens array is spatial variant, we can segment the object domain into a grid of small zones, within which the PSF is approximately spatial invariant in lateral direction and scales linearly with the object depth (Fig. 2). Here, we scale and project the segmentation lines between the individual microlens units onto a reference object plane. This segments the reference object plane into $N$ zones, with $(x_{ci}, y_{ci})$ being the centroid of each zone, where $i = 1, 2 \ldots N$ and $N$ is the number of microlens units. Defining $f(x, y; z)$ as the object intensity, $u_i$ as a 2D circular boundary operator to select the objects in the $i^{th}$ zone, and $\odot$ as an element-wise multiplication, the local object falling into the $i^{th}$ zone can be written as $f(x, y; z) \odot u_i(x - x_{ci}, y - y_{ci})$ (Fig. 2A).

The PSF can be derived from the arrangement of the microlens units. We define $h_0(x', y') = \sum_i \delta(x' = x'_{ci}, y' = y'_{ci})$ on the image plane, where $(x'_{ci}, y'_{ci})$ is the centroid of each microlens unit. Any object point can only be effectively imaged by a certain amount of lens units $N_e$ (across $\overline{AB}$ in Fig. 2A), onto an effective area $A_e$ (across $\overline{CD}$ in Fig. 2A) on the image sensor, due to the Lambert's cosine law [21] of light intensity distribution falloff and geometric aberration of the lens unit. Therefore, we can use an envelope function $C(x', y')$ to model the overall envelope of the PSFs (Fig. 2B). The local PSF for the $i^{th}$ object zone can then be expressed as $\boldsymbol{\mathcal{D}}_z[h_0(x', y') * PSF_o(x', y') \odot C(x' - x'_{ci}, y' - y'_{ci})]$, where $PSF_o(x', y')$ is the PSF of an individual lens unit, $*$ represents convolution, and $\boldsymbol{\mathcal{D}}_z$ is a local scaling operator to scale the coordinates (with a factor of $s_z$) to obtain the local PSF for different object depth $z$. $s_z$ can be expressed as

$$s_z = \frac{\overline{CD}}{\overline{AB}} \cong \frac{z + v}{z} \tag{1}$$

The image on the image sensor $b(x', y')$ can be expressed as the summation of local image from all zones and object depths. The local image in a certain zone and object depth is the convolution between the local objects and local PSFs, with an appropriate lateral magnification $M_z$ of the microlens unit (see section Object Reconstruction for a further discussion on $M_z$).

$$b(x', y') = \sum_z \sum_{i=1}^{N} \left[ f\left(\frac{x'}{M_z} + x_{ci}, \frac{y'}{M_z} + y_{ci}; z\right) \odot u_i\left(\frac{x'}{M_z}, \frac{y'}{M_z}\right) \right] * \boldsymbol{\mathcal{D}}_z[h_o(x', y') * PSF_o(x', y') \odot C(x' - x'_{ci}, y' - y'_{ci})] \tag{2}$$

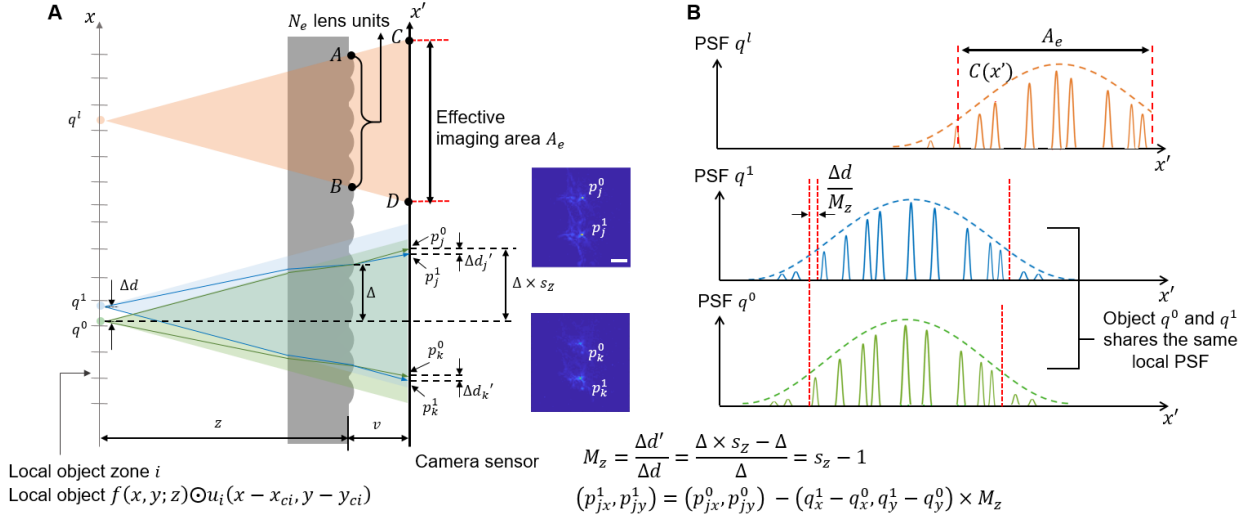

**Fig. 2. Image formation and object reconstruction.** (A) Each point source is effectively imaged by $N_e$ lens units, and forms an imaging area $A_e$ on the image sensor. For each object point $q$ to be reconstructed, all the imaged points $p$ on the image sensor are found within the effective imaging area $A_e$ through the local PSF and the lateral magnification $M_z$. Scale bar, 60 μm for $\Delta d$=952 μm. (B) Local PSF for the individual point sources. The intensity falloff is characterized by the envelop function $C(x',y')$, attributed to the angular distribution of the point source intensity and the geometric aberration of the lens units. As $q^0$ and $q^1$ are two object points in the same local zone, they share the same local PSF.

**Object reconstruction**

So far, nearly all the lensless imagers employ iterative optimization approach with sparsity constrains to reconstruct the objects. To overcome the challenge of high demand of computation resource, we developed a pixel back projection algorithm followed by background suppression. The required computational resource is greatly reduced, and scales linearly with the number of reconstruction voxel in the object.

The pixel back projection algorithm uses geometrical optics approach to reconstruct the object at different depths (Fig. 2A), which essentially solves the inverse process of Eq. 2. As each object point projects to multiple points on the camera sensor through different lens units, our goal is to collect these points on the camera to recover the object points. While each lens unit could image a point source into multiple pixels on the camera sensor (defined by $PSF_0$), to simplify the process, we only pick up the centroid pixel, as the others are typically much weaker (Fig. 2A). We start with a calibration process of $\mathcal{D}_z[h_0(x',y')]$, by finding the scaling factor $s_z$ between the local PSF for object depth $z$ and the microlens arrangement pattern $h_0(x',y')$. Ideally, $s_z$ is a constant across the field of view at the same object depth, and it scales linearly with depth. Thus, we only need to measure $s_z$ at one object point location. In practice, we perform the measurement by placing a point source across multiple reference points in the object space, so as to align the microlens array with the camera sensor, and to capture any deviation in the $h_0(x',y')$ of the fabricated microlens array from design. The magnification $M_z$ of each lens unit can be approximated through the scaling factor $s_z$ (Fig. 2A):

$$M_z \cong s_z - 1 \qquad (3)$$

Once we have $\mathcal{D}_z[h_0(x',y')]$ and $M_z$, we can locate all the pixels for a voxel of interest $p$ at depth $z$. The reconstructed voxel value is the sum of all mapped pixel values across the effective image area $A_e$, which can be determined based on the field of view of the single lens unit (Materials and Methods). Such a reconstruction strategy avoids considering all pixels in the image sensor to reconstruct a single object voxel, which is the key difference from the global iterative optimization approach. It is this local reconstruction strategy that significantly reduces our computation resource.

The reconstructed results in each depth from pixel back projection contain both the in-focus objects and artifacts, which includes the defocused light from objects locating at other depths and possible ghost objects. The ghost objects originate from the "one-to-many mapping" nature of microlens array so a single pixel on the camera could come from multiple object points. All these contribute to the background and becomes more severe when the object is less sparse. The second step of our reconstruction process is to suppress these backgrounds, by either a particle clustering algorithm [22] or a U-net based convolutional neural network [23].

For sparse point objects, we developed a particle clustering algorithm to remove the ghost object points and the defocused light in the 3D stack reconstructed from the pixel back projection algorithm (Materials and Methods, Supplementary Section S1, Fig. S1-2). The clustering algorithm is based on graph connectivity. For less sparse 3D objects, we developed a convolutional neural network to suppress the background (Materials and Methods, Supplementary Section S2, Fig. S1 and S3). It slices overlapped objects in depth and picks out focused objects from background light. The output image from the neural network contains sharper contrast and the defocused features are largely removed.

**Design of microlens array**

We designed a compact microscope that can image dense fluorescent objects (~520 nm central wavelength) across a large volume (~23×23 mm$^2$ field of view and ~5 mm depth of field) with a lateral resolution ~40 μm. The microlens array and the image sensor both have a size ~20×20 mm$^2$, matching the field of view. To reduce the thickness of the microscope, we minimize the distance between the microlens array and the image sensor to be <5 mm. For a convenient placement of samples, we set the nominal working distance as 20~30 mm. Based on this performance metrics, we can optimize the design parameters of the microlens array (Materials and Methods, Supplementary Section 3, Fig. S4-6). For a single lens unit, our major optimization interests are depth of field, lateral resolutions and aberrations. For the performance of the entire lens array, our optimization focuses on balancing 3D resolving ability and image reconstruction speed. We summarize the design parameters and performance metrics in Table 1.

| Design parameters | | | | Performance metrics | |
|---|---|---|---|---|---|
| Working distance $z$ | 18~30 mm | MLA-camera distance $v$ | ~4 mm | Lateral resolution | ~40 μm |
| Lens array area | 20×20 mm$^2$ | Number of lenses | 200 | Axial resolution | ~300 μm |
| Average pitch size | 1.23 mm | Lens focal length $f$ | ~4.65 mm | Field of view | ~23×23 mm$^2$ |
| Lens refractive index | 1.43 | Lens radius of curvature | 2 mm | Depth of field | ~5 mm |
| Effective imaging lens number for single point object $N_e$ | 15~20 | Diameter of the effective field of view of single lens unit $2\sqrt{A_e/\pi}$ | ~ 6 mm | Magnification | 0.1-0.15 |
| Camera pixel size $d_p$ | 4.5×4.5 μm$^2$ | Emission wavelength | 500-530 nm | Sensor occupancy parameter $v = N_e \times M_z^2$ | ~0.4 |

**Table 1. Design parameters and the derived/simulated performance of GEOMScope.**

**Simulation results**

To verify the design of GEOMScope, we first simulated its ability to image a single point source. We used ray tracing algorithm (OpticsStudio) to form an image of a single point source (Fig. 3). Reconstruction of the object shows a FWHM ~40 μm and ~300 μm in the lateral and axial direction respectively (Fig. 3A-B). We then simulated its ability to resolve two point sources separated by 40 μm at a depth of 20 mm. The two point sources could be clearly distinguished in the reconstructed object (Fig. 3C), where the intensity drops below ~85% of the peak object intensity in between the two reconstructed points. Similarly, two points sources that are separated by 300 μm along the same axial axis could be distinguished in the reconstruction (Fig. 3D). These simulation results agree with our design.

We then simulated the image formation and object reconstruction for 3D distributed fluorescent point sources (Fig. 4A) and a 3D volumetric object (Fig. 4B), which are placed at a distance of 18~22 mm from the front surface of lens array. The images are again formed by ray-tracing algorithm. The pixel back projection algorithm reconstructs the objects at different depths. These reconstructed object stacks are then fed to the particle clustering algorithm (Fig. 4A) or convolutional neural network (Fig. 4B) for background suppression. The objects in 3D can then be faithfully reconstructed.

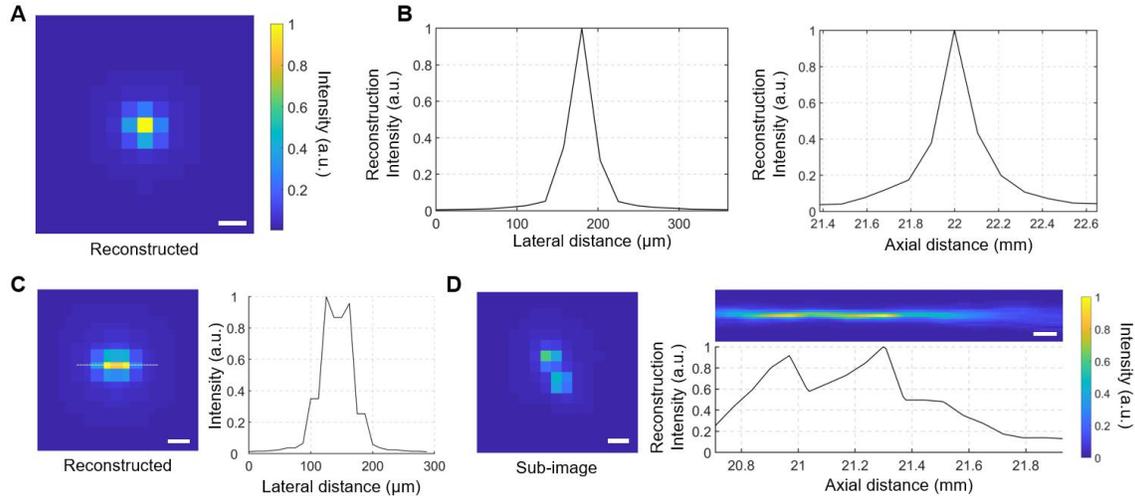

**Fig. 3. Simulation on the 3D resolving power of GEOMScope.** (A) The lateral spatial profile of the reconstruction of a single point object. Scale bar: 40 µm. (B) Line profile of the reconstructed single point object along the lateral (left) and axial (right) direction. (C) Simulated two-point lateral resolution: two object point sources laterally separated by 40 µm at z=20 mm can be resolved. Left, the spatial profile of the reconstructed object; right, line profile along the lateral direction of the reconstructed object. Scale bar: 40 µm. (D) Simulated two-point axial resolution: two object point sources axially separated by 300 µm at z=20 mm can be resolved. Left, the sub image from one lens unit, showing the two axially separated point sources are separated laterally in the sub image. Right, the reconstructed objects with its line profile along the axial direction. Scale bar: left, 10 µm; right, 100 µm.

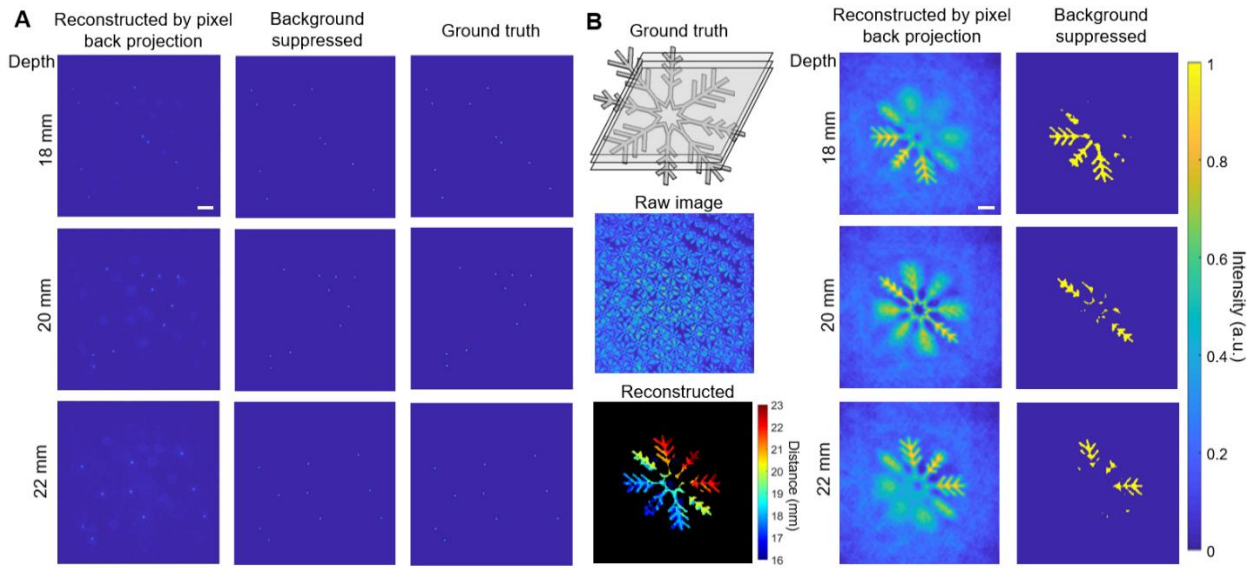

**Fig. 4. Simulated image formation and reconstruction of fluorescent objects in 3D.** (A) Imaging and reconstruction of 3D distributed fluorescent particles, for a depth of 18 mm, 20 mm, 22 mm. Left column: reconstructed objects through pixel back projection only; middle, final reconstructed objects after background suppression through particle clustering algorithm; right, ground truth object. The point objects are smoothed on intensity profile to make their positions more visible. Scale bar, 2 mm. (B) Imaging and reconstruction of a 3D snowflake fluorescent object, for a depth of 18 mm, 20 mm, 22 mm. Left column: top, 3D ground truth object; middle, raw image; bottom, reconstructed volumetric object. Middle column, reconstructed objects through pixel back projection only; right column, final reconstructed objects after background suppression through convolutional neural network. Scale bar: 2 mm.

**Experimental imaging of point source and resolution target**

We fabricated the microlens array using optical transparent polydimethylsiloxane (PDMS) [SYLGARD® 184] with a negative 3D printed mold (Materials and Methods, Fig. S7). To quantify the image quality of GEOMScope across the 3D volume, we measured the FWHM of the lateral and axial reconstruction profile of a point source (created by illuminating a 10 µm pinhole) located at different positions in the object space. Here, we removed the filter so the imaging wavelength is ~457±25 nm. The lateral and axial reconstruction profile of the point source maintains stable for a depth range over ~5 mm when it is placed ~20 mm from the lens array (Fig. 5), in an overall good agreement with the design (Supplementary Section 4).

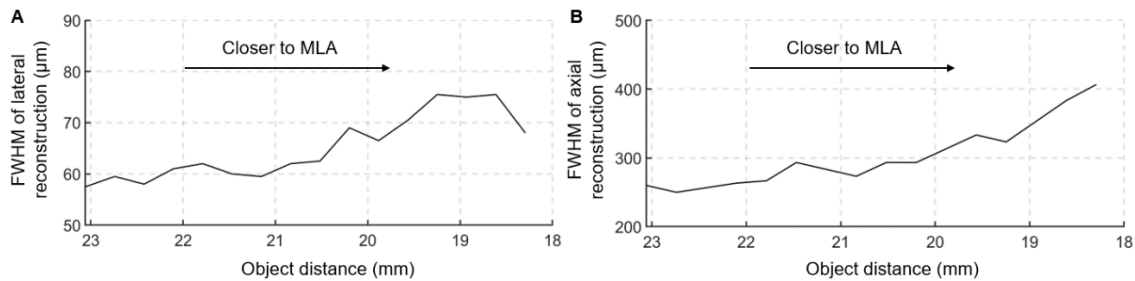

**Fig. 5. Experimental measurement of the reconstruction profile of a single object point versus object depth.** (A) FWHM of the point reconstruction profile along lateral direction. (B) FWHM of the point reconstruction profile along axial direction.

To further quantify the imaging resolution, we imaged a negative 1951 USAF resolution target. We inserted a diffuser (600 Grit) between the LED and the resolution target to increase the angular spread of light passing through the target, so that the target acted like an isotropic source as a fluorescence sample. We placed the target at different depth, captured single-shot images, and performed image reconstruction in 3D for each case (Fig. 6). For 2D objects, the pixel back projection step alone can already achieve a high quality reconstruction. The reconstruction at different depths clearly demonstrated the effectiveness of our algorithm: the objects reconstructed at the correct focal plane show the highest sharpness. We can clearly resolve group 3 element 2, which has a line space of 55 μm (Fig. 6B). This is in a reasonable agreement with the simulated two-point resolution.

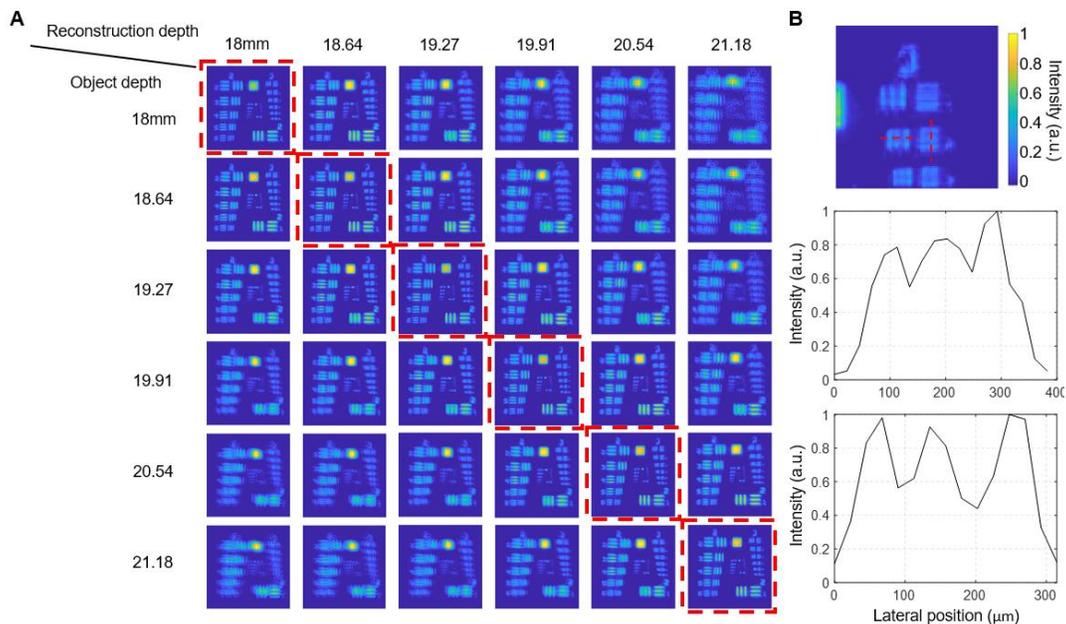

**Fig. 6. Imaging and reconstruction of 1951 USAF resolution target for group 2 and 3.** (a) Object reconstruction at various depths, when the target was positioned at a depth of 18 mm, 18.64 mm, 19.27 mm, 19.91 mm, 20.54 mm, and 21.18 mm. The best reconstruction corresponding to the actual target depth is labeled by a red dashed boundary. No background suppression was employed for this 2D sample. (b) Zoom-in view of the best reconstruction of target at 18 mm depth, and the line profiles of the horizontal line pairs and vertical line pairs of group 3, element 2.

**Experimental imaging of 3D fluorescence samples**
We fabricated a 3D fluorescent sample by embedding fluorescent beads [Firefli Fluorescent Green (468/508nm), 5.0 μm, Thermo Scientific] into SYLGARD 184 PDMS. The sample is 35 mm in diameter and ~2.5 mm thick, where clusters of fluorescent beads were randomly distributed. We captured the fluorescence image through GEOMScope, and a benchtop inverted microscope with a 1x objective lens as a control (Fig. 7). The 1x objective lens provides a large field of view, but at a tradeoff of a poor axial resolution in the benchtop microscope. Nonetheless, it requires multiple exposures and subsequent image stitching to cover the entire field of view (23mm×23mm). The depth information is lost due to the poor axial resolution. In comparison, only a single exposure is needed in GEOMScope to cover the entire field of view, and the bead clusters at different depths can be resolved. This clearly demonstrates the strength of GEOMScope: the ability to capture 3D images across a large field of view.

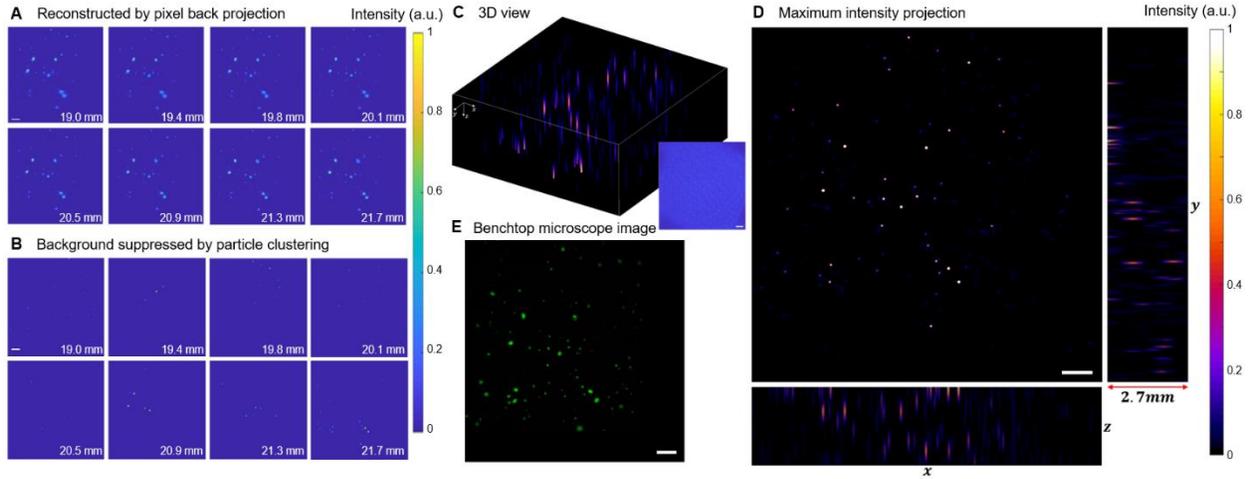

**Fig. 7. Imaging of 3D distributed fluorescent particles.** (A) Reconstruction from pixel back projection algorithm for different depth. (B) Background suppression of (A) through the particle clustering algorithm. Different particle clusters are separated at different depths. The clustered particles are intentionally blurred for a clearer visualization. (C) 3D volumetric view of the reconstructed volume. Each point has a fixed axial span, the same as the depth interval between adjacent slice in the reconstruction. Inset: raw captured image. (D) Maximum intensity projection view of the reconstructed particle clusters in 3D at xy, xz and yz plane. (E) Images captured by a benchtop microscope and stitched for same field of view as (C-D). Scale bar for (A-E): 2 mm.

We next tested the GEOMScope on resolving large scale 3D volumetric object, which is not sparse in spatial domain (Fig. 8). We 3D printed a snowflake mold with feature thickness 0.7 mm on a clear substrate (Proto Labs, Inc.). Glowing powders in green color was mixed with PDMS and spread onto the mold to form phosphorescent features. The target was tilted with respect to the lens array surface and spanned a depth range of 20~25 mm from microlens array. While the object was relatively dense so different sub-images overlapped (Fig. 8A), the reconstruction algorithm successfully recovered the 3D information (Fig. 8B-C, E). There are a total number of $\sim 4 \times 10^6$ voxels and $\sim 10^{14}$ pixel-voxels pairs in this 3D reconstruction. The processing time is ~18 second in total using a workstation (8 threads parallelization).

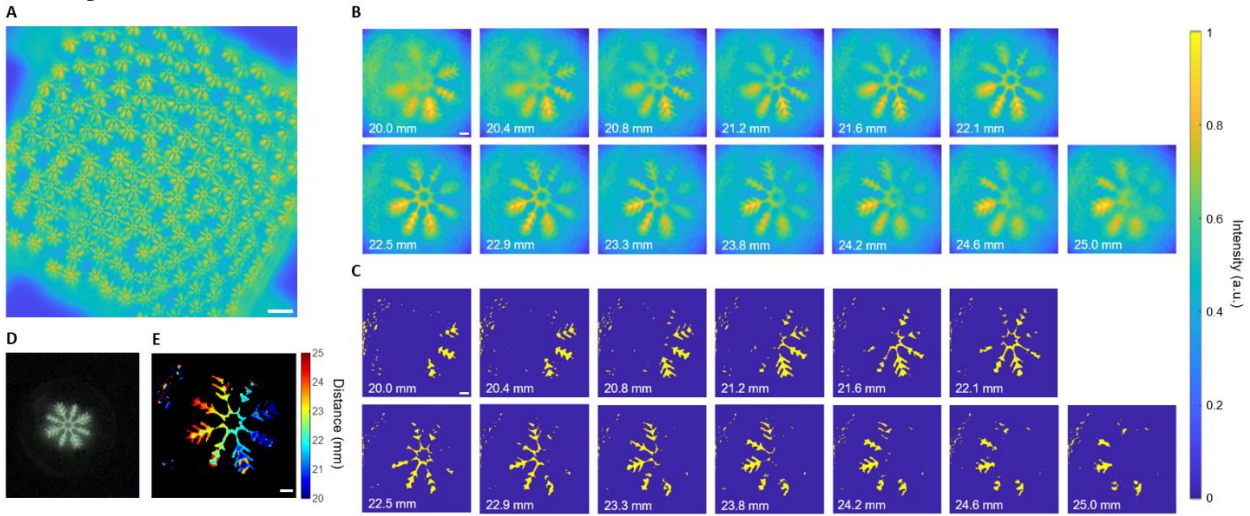

**Fig. 8. 3D imaging of large scale phosphorescent objects.** (A) Raw image. (B) Reconstruction from pixel back projection algorithm for different depth. (C) Background suppression of (B) through a convolutional neural network. As the feature thickness is 0.7 mm in the object, the same feature could appear in multiple depth slices in the reconstruction. (D) Macroscopic photo of phosphorescent object. (E) Reconstructed depth-resolved volumetric object. Scale bar for (A-C), (E): 2 mm.

**Computation speed of pixel back projection and comparison with iterative optimization algorithms**

A key advantage of GEOMScope is that the object reconstruction has a much lower computation cost and achieves a much faster reconstruction than the existing algorithms. To quantitatively validate this assessment, we compared the object reconstruction time using pixel back projection and the iterative optimization algorithms such as ADMM or Richardson–Lucy deconvolution, for small data scales with pixel (camera)-voxel (object) pair number from $\sim 10^6$ to $\sim 2 \times 10^9$ on a workstation (Intel Xeon E5-2686 v4, 128 GB RAM, MATLAB 2019b) [Figure 9(A), see also Supplementary Table S1]. The objects are distributed particle sources on a 2D plane. For data size within the physical RAM limit, the computation time increases much faster with the number of possible pixel-voxel pairs when using optimization solvers and Richardson–Lucy (R-L) deconvolutions, while pixel back projection method has a relatively stable computation time on the order of milliseconds. The ADMM solver exceeds physical RAM limit when the number of pairs reach $\sim 10^9$, resulting in a rapidly increased computing time. The process time of both R-L deconvolution and ADMM solvers increase linearly with the number of pixel-voxel pair. This is expected as the number of elements in the system matrix in image formation (and thus the number of scalar multiplication in matrix operation) equals to (scales with) the number of pixel-voxel pair. In pixel back projection, for each voxel, we only pick one pixel in one sub-image of the lens unit within the effective image area, so the computation time only scales with the number of reconstructed voxel. As we set the same number of pixel and voxel for each data point in Fig. 9(A), the computation time appears to scale sublinearly (with a theoretical slope of 0.5) with the pixel-voxel pair. In fact, the required reconstruction time in GEOMScope is so small by itself at this range of data scale that the total processing time is dominated by the program initialization, resulting in a slope of <0.5 between the computation time and pixel-voxel pair number. We note that the increase of pixel number in the image could increase the resolution and thus the object complexity that the reconstruction algorithm can handle. However, it will not increase the processing time, which is one advantage of our algorithm. In terms of RAM usage, deconvolution and optimization solver requires additional RAM to store the entire system matrix and operate the matrix algebra, while our method only requires a minimum amount of RAM as only one pixel-voxel pair is traced at each step. Our method using geometrical optical prior from microlens arrangement is thus much more suitable to solve large scale image reconstruction problems, though it suffers from a poorer reconstruction quality resulted from ghost objects. The reconstruction quality is greatly improved with the background suppression algorithm and becomes similar as those using ADMM solver or Richardson-Lucy deconvolution when the number of pixel-voxel pair increases [Fig. 9 (B-D); Supplementary Section 5, Fig. S8].

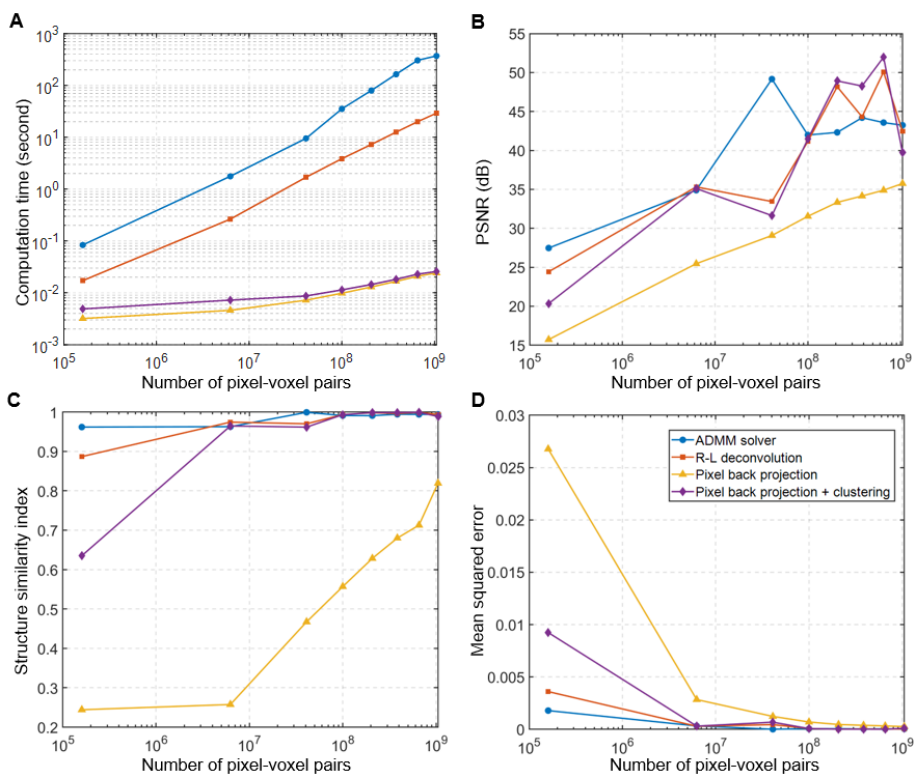

**Fig. 9. Comparison between pixel back projection method and iterative-optimization-based methods on computational cost and reconstruction quality merit.** (A) Computational cost on a workstation (Intel Xeon E5-2686 v4, 128 GB RAM, MATLAB 2019b). (B) Peak signal-to-noise ratio (PSNR) of the reconstruction. (C) Structure similarity index (SSIM) of the reconstruction. (D) Mean squared error of the reconstruction. (B)-(D) is compared to ground truth. The computation cost of the two optimization-based methods, ADMM and Richard-Lucy deconvolution increases rapidly with data scale, while it only increases modestly in our pixel back projection method. When paring with the particle clustering algorithm for background suppression, the pixel back projection method has a greatly improved reconstruction quality, and approaches those from the two optimization-based methods when the number of pixel-voxel pairs increases.

Benefited from its local reconstruction nature, the speed of the pixel back projection algorithm can be further increased by parallelizing the pixel-voxel ray matching in reconstruction. Multi-thread processing can be applied to resolve objects in each local region across different depths while maintaining the overall reconstruction rate. We further analyzed the efficiency of the parallel computing with multi-threads by assigning different reconstruction depths to individual threads, and compared the reconstruction speed to that using single thread. In our tests, the effectiveness of parallel computing is significant when the data size or number of voxel-pixel pairs is larger than $2 \times 10^7$. Compared with parallelized optimization process using gradient descent, our multi-threading method is straightforward and simple to conduct. For 13 depths reconstruction in our method (~$10^{14}$ voxel-pixel pairs), parallel computing with 4 threads on a desktop yields a speed increase of 230%, with a total computing time 30 seconds; parallel computing on a workstation with 8 threads yields a speed increase of 380% and a total computing time 18 seconds. Through further optimization on data structure and more threads, we expect reconstruction of large 3D volume with speed close to video frame rate.

**DISCUSSION**

We demonstrated a lensless integrated microscope that can perform 3D imaging across a large field of view with high resolution and fast object reconstruction speed (Fig. 1). The pixel back projection algorithm, in combination with efficient background suppression algorithm, significantly reduces the computational cost and increases the object reconstruction speed, while preserving a good resolution. It enables large field-of-view microscopic imaging with mega pixel reconstruction of the object, which would consume an unrealistic amount of computation resource and processing time using the prevailing reconstruction algorithms through iterative optimizations or the more recently proposed deep neural network. We designed our microlens array based on the performance metrics of individual lens unit as well as lens array in both geometrical and wave optics aspects. We demonstrated the performance of our methods in theoretical modeling (Table. 1), numerical simulation (Fig. 3-4) and experiment (Fig. 5-8), and showed excellent results across different types of samples ranging from point source (Fig. 5), resolution target (Fig. 6), fluorescent particles (Fig. 7) and volumetric objects (Fig. 8). We also verified the efficiency of our reconstruction algorithm by a quantitative comparison on the reconstruction speed with the iterative optimization approaches (Fig. 9). This work presents a promising approach of applying miniaturized lensless imaging device for high speed, high throughput imaging.

Compared to conventional cameras and microscopes where there is a one-to-one pixel mapping between the object plane and the sensor through the bulk lenses, lensless imagers map a single voxel in 3D objects to multiple pixels on a 2D camera through a thin piece of optics [1, 2, 4-7, 24-26]. In general, the imaging system can be formulated as $\boldsymbol{b} = \boldsymbol{A}\boldsymbol{f} + \boldsymbol{g}$, where $\boldsymbol{f}$ is the object, $\boldsymbol{A}$ is a linear measurement matrix containing the PSF of each point in the 3D object space, $\boldsymbol{b}$ is the measured data on the camera, and $\boldsymbol{g}$ represents the noise. Depending on the imaging optics, which fundamentally determines $\boldsymbol{A}$, there have been two major classes of lensless cameras reported (Supplementary Table 1). The first class is amplitude mask [2, 6, 27-29]. In Ref. [2, 6], the amplitude mask was designed such that the measurement matrix $\boldsymbol{A}$ could be broken down to two small matrices, which greatly reduces the required memory and improves speed in object reconstruction. A major drawback of this approach is the low light throughput and high sensitivity to alignment between the mask and the camera. The second class, which employs a phase mask such as a diffuser [1, 30] or a microlens array [4, 5, 7, 31, 32], increases the light throughput. Diffusers have a dense PSF requiring large spatial support, and it is not suitable for pixel back projection. When using a microlens array, the PSF (and thus $\boldsymbol{A}$) is relatively sparse. This could increase the signal-to-background ratio in the images and thus relax the general requirement of sparse objects in lensless imaging. We thus choose a random layer of microlens array as the imaging optics. The random position of lens units reduces the crosstalk between lenses, which facilities the pixel back projection algorithm.

Our method solves the general challenges of lensless imagers where there is a high demand of computation resources for object reconstruction. In the prevailing iterative optimization algorithms, the object is solved by mapping the entire object world to the camera sensor and performing a global optimization. Typically, the $L_1$ norm regularized total variation is used in the optimization problems, as sparse 3D gradients work well in most imaging applications. In addition, both $L_1$ and $L_2$ regularization terms may be adopted to resolve natively sparse objects or reduce noise amplification [4-6, 33]. While this could typically achieve a reconstruction result in high fidelity, the global mapping could contain high redundancy when an object feature is only effectively imaged by a local part of phase mask. In addition, it requires complex operation on a large matrix with the number of matrix elements being the product of camera pixel count and object voxel count, leading to a very high computation cost. This restrains the application in high resolution large field of view imaging. In GEOMScope, we use a random microlens array as the imaging optics, which has a relatively sparse PSF and a near identity transformation matrix between the object and each sub-image. This allows us to use the pixel back projection algorithm, which reconstructs each object voxel by summing the corresponding camera pixels according to the geometry of the microlens array. Each object voxel is reconstructed from only those lens units whose effective imaging area covers the voxel. This essentially turns the global optimization into local reconstruction, where different voxel reconstruction can run independently. Such approach shows excellent scalability with the data size, and is very suitable for high resolution large field of view imaging. Given the fixed lens unit size and object-lens-camera distance, the field of view can be increased by increasing the number of lens unit. Thanks to the local reconstruction mechanism, such an increase of field of view does not increase the reconstruction complexity of each object voxel. In GEOMScope, we pair the large-area microlens array with a high-pixel-count large-sensor-area camera. This enables a high resolution large field of view imaging, with high counts of object voxels, and fast reconstruction speed, which would otherwise consume a very expensive computation resource using global optimization algorithm. Furthermore, the local reconstruction nature of our algorithm works very well with parallelized processing. Our computation imaging strategy combining a sparse PSF design and localized reconstruction through pixel back projection is thus an effective way for large field of view lensless microscopic imaging.

We note that GEOMScope has a high reconstruction resolution. The PSF of each lens unit is well confined within the 3D resolving range, and this ensures a high efficacy of reconstruction using geometric optics. We achieve a resolution close to that determined by wave optics. This is very different from the early generation of light field camera [34] or light field microscopy [18]. There, the PSF is not locally sparse, so reconstruction through geometric optics suffers from a large uncertainty, bounded by the size of the microlens unit. Iterative optimization approach takes full information of the complex PSF and can thus achieve a much better resolution. In GEOMScope, we configure the microlens array and the system magnification such that the PSF behaves as a Dirac comb function. This results in a highly precise image formation modeled by geometric optics, and thus a high resolution reconstruction without using iterative optimization approach.

GEOMScope shares some similarity with focused plenoptic camera [20] as the PSF in both is well confined. Though both of their reconstruction algorithms are based on geometric optics, they are fundamentally different. The reconstruction algorithm in the focused plenoptic camera is designed for photography application in macroscale, and would fail for microscopy application. There, the object is reconstructed by selecting sub-images from each microlens and stitching them together. The sub-image patches are selected based on the depth of the spatial scene estimated by local image correlation between neighboring microlens units [20]. Each region of the reconstructed object only uses one patch from a single sub-image. For microscopic imaging, the system magnification becomes much larger, leading to a substantial crosstalk between the sub-images of microlens units. Using image-patches to fill up the reconstruction space will thus generate a lot of artifacts. In contrast, the pixel back projection algorithm used in GEOMScope reconstructs each voxel by collecting a single image pixel from each sub-image and summing them across multiple sub-images. Such strategy makes it highly effective to handle crosstalk and generate high quality reconstruction.

Another feature of our work is the innovative implementation of background suppression after the pixel back projection algorithm. Unlike global optimization approach, background such as the out-of-focus light or ghost object is inevitable in pixel back projection as it performs local reconstruction without consideration of other object source points. An additional background suppression step is thus critical to enhance the overall reconstruction performance. It allows the suppression of out-of-focus light as well as the ghost objects, and brings the reconstruction quality similar to those of iterative optimization algorithms. We note that this extra step does not increase the computation time too much, particularly for the particle clustering algorithm. The convolutional neural network requires a bit more computation resource (33 ms inference time for each object slice, Intel Xeon CPU @2.30GHz, NVIDIA Tesla T4 GPU), but still significantly lower than the iterative optimization algorithms or the deep neural network directly mapping the images to the objects. For sparse samples with featureless characteristics, we adopt particle clustering algorithm because of its simplicity and fast processing time. For complex 3D object, convolutional neural network is used. Comparing with other object detection methods such as circle detection through Hough transform [35] and bag of words feature recognition [36], convolutional neural network has a higher precision to select the in-focus objects, a better scalability to denser and more overlapped objects, and a better generalization to object shapes and sizes. In our current implementation, so as to reduce the required computation resource, the neural network takes a downsampled, single layer (depth) image as the input, instead of a full resolution 3D stack. We note that the downsampling reduces the resolution, and single layer input could result in a less desired reconstruction quality when two objects with very different intensity levels are close in depth. In the future work, we aim to enhance the performance of neural network by including the physical mechanism of pixel back projection to form trainable inversion module of the network for better flexibility, and training it with multiple layer inputs without downsampling of the dataset. Furthermore, by combining the loss function with regularization terms and total variation, it may be generally applied to sparse objects with a larger dynamic range of intensity levels.

One potential limitation of GEOMScope is that the geometrical design of the microlens array is application-driven; thus one configuration of microlens array fit for a specific application scenario may not be suitable for another. For a specific design, as the object density increases beyond a specific value such that sub images of different lens units overlap significantly, the signal-to-background ratio of the reconstruction would degrade, even with background suppression algorithm. This is in fact a similar challenge encountered by the global optimization algorithms, though there is some flexibility in redesigning the regularization term (e.g. replacing $L_1$ norm of object into $L_2$ norm or 3DTV) to alleviate this issue. In GEOMScope, reconfiguration of lens unit sizes/density and system magnification is necessary to control $N_e$ and $A_e$ to maintain a good 3D resolving ability. We note that the sensitivity to object sparsity is a general challenge for lensless imaging and all reconstruction algorithms, and is an on-going research topic.

In the current version of GEOMScope, the resolution and thus the number of resolved voxels are limited by the system magnification (Materials and Methods). We aim for a large field of view imaging and thus choose a camera with large sensor area. However, the mechanical casting poses a minimum distance between the microlens array and the camera sensor. This reduces the magnification and thus limits the resolution and the number of resolved voxels. Nevertheless, we achieve a resolution close to that dictated by wave optics. By further engineering the camera and optimizing the microlens design, GEOMScope could achieve a higher resolution and more resolved voxels.

## MATERIALS AND METHODS
### Design consideration of the GEOMScope
We choose the design parameters of GEOMScope, including the pitch, focal length, total size of the microlens array, as well as object-microlens-array distance, and microlens-array-camera distance, based on the performance metrics.

*Lateral resolution*

The lateral resolution of GEOMScope depends on the Abbe diffraction limit of a single lens unit. At light wavelength $\lambda$, the minimum resolvable distance of image spots on sensor is $\lambda/(2NA)$, where $NA$ represents the numerical aperture of a single lens unit. The theoretical diffraction limited lateral resolution $r_{xy0}$ can be expressed as

$$r_{xy0} = \frac{\lambda}{2NA} \times \frac{1}{M_z} \tag{4}$$

Considering the finite pixel size $d_p$ on the camera sensor, the final lateral resolution is expressed as

$$r_{xy} = max\left(\frac{\lambda}{2NA}, d_p\right) \times \frac{1}{M_z} \tag{5}$$

*Axial resolution*

The object movement in the axial direction results in a lateral shift of its images on the sensor. We can thus model the axial resolution $r_z$ as the distance change of an object point that leads to a shift of the image point at the boundary of the effective sensing area $A_e$ in a distinguishable value $r_{xy}$ (Fig. S4):

$$r_z = r_{xy} \times \frac{z}{\sqrt{A_e/\pi}} \tag{6}$$

where $\sqrt{A_e/\pi}$ is the radius of effective imaging area on sensor. Here, we assume the axial movement is small compared to the object distance, thus the variation of lateral resolution can be ignored.

*Depth of field*

For small object-lens distance $z$ and low numerical aperture, the depth of field is determined by the confusion circle of each lens. When the object moves away from the focal plane, the imaged spot on the camera sensor will spread out in the lateral direction, forming a confusion circle. We set the diameter of the confusion circle to be two-pixel size on the sensor, and the corresponding DOF can be calculated following the geometrical optics [37] (Supplementary Section S3, Fig. S5):

$$DOF = z_F - z_N \cong \frac{2F_\# cz^2 f^2}{(f^4 - F_\#^2 c^2 z^2)} \tag{7}$$

$$z_N = \frac{zf^2}{f^2 - cF_\# f + cF_\# z}, \qquad z_F = \frac{zf^2}{f^2 + cF_\# f - cF_\# z} \tag{8}$$

where $F_\# = f/D$ is the f-number of the individual lens unit, $f$ and $D$ being the focal length and diameter of the lens units respectively.

*Field of view*

The field of view is measured as the lateral range of objects within the depth of field distance range that can be effectively imaged onto the camera sensor. Unlike the conventional microscope where the field of view is much smaller than the imaging optics, GEOMScope can have a field of view in a similar size as the microlens array.

One related concept is the field of view of an individual lens. A larger field of view of a single lens unit is equivalent to a larger $N_e$, i.e. more lens unit could image a single object point, thus increasing the accuracy of reconstruction through pixel back projection. The field of view of a single lens unit can be evaluated when the Strehl ratio of the PSF falls below a threshold. The Strehl ratio $Strehl$ can be approximated through the RMS wavefront error $\Delta w$ from coma ($\Delta w_c$), and spherical aberration ($\Delta w_s + \Delta w_d$) [38]:

$$Strehl \cong exp[-(2\pi\Delta w)^2] \tag{9}$$
$$\Delta w = \Delta w_c + \Delta w_s + \Delta w_d \tag{10}$$

In general, a smaller lens diameter $D$ and a larger focal length $f$ would reduce the aberration and thus result in a larger field of view of a single lens unit.

*Consideration of object sparsity*

As one object point source is imaged by $N_e$ microlens units, some level of sparsity is required on the object so as to reconstruct objects effectively without regularization. If there is a substantial overlap between the sub image from each lens unit, the reconstruction becomes ill-posed. From the perspective of pixel projection, we define an occupancy parameter $V$ to describe the percentage of the camera sensor area being illuminated when an object occupies the entire field of view at a single depth:

$$V = N_e \times M_z^2 \tag{11}$$

In general, a small $V$ allows denser objects to be imaged and reconstructed.

*Design of the microlens array*

We aim to design GEOMScope such that it can image fluorescent objects (~520 nm central wavelength) across a large volume ~23×23×5 mm³ with a lateral resolution ~40 μm. The distance between the microlens array and the image sensor is <5 mm whereas the nominal working distance is 20~30 mm. We choose an off-the-shelf camera (DALSA Genie Nano-CL M5100 NIR, Teledyne) equipped with PYTHON25K CMOS image sensor (ON Semiconductor) as it has a large sensor area (23×23 mm²), pixel count (5120×5120), and a small pixel size of 4.5×4.5 μm², suitable for large field-of-view high resolution imaging.

The working distance and the object-microlens-array distance determines the magnification $M_z$~0.15 for each lens unit. To resolve objects 20~30 mm away from the microlens array with ~5 mm depth of field (radius of confusion circle ~$2d_p$), the numerical aperture of the individual lens unit should be less than $NA_{max} \cong 0.20$ (Eq. 7~8) when we set a ~4 mm focal length. A larger numerical aperture is allowed when we reduce the focal length. Meanwhile, the lateral resolution and the magnification sets a lower limit of the numerical aperture $NA_{min} \cong 0.045$ (Eq. 5). The number of lens on the lens array could impact the computation cost in object reconstruction (pixel back projection), the occupancy parameter (Eq. 11), and the 3D resolving capability. On the one hand, to reduce the computation cost in object reconstruction and the occupancy parameter, the total number of lens should be small. On the other hand, more lens units to image an object point could provide more angular information of the light rays and thus improve the 3D resolving ability. Considering all the above factors, we set the diameter $D$ and the focal length $f$ of an individual lens unit as 2 mm and 4.65 mm respectively. ~200 lens units are randomly positioned across the 20×20 mm² lens array area, with a fill factor ~1. The partial overlap between the lenses results in a reduced pitch size to ~1.23 mm on average, leading to an effective NA of ~0.13 for each lens. We simulate the aberration of the lens unit (Supplementary Section 4, Fig. S6). Setting a Strehl ratio above half of its peak value, we determine that the diameter of the effective field of view of each lens unit is ~6 mm, and one object point can be imaged by $N_e$~20 lens units (corresponding to an effective image area ~9π mm²). With $M_z$~0.15, we have the ratio between total area of sub images and sensor area ~0.4 (Eq. 11). This allows imaging dense 3D objects. The design parameters and the calculated performance metrics are summarized in Table 1.

**Fabrication of microlens array and setup of GEOMScope**

Based on the design parameters, we generated a 3D layout of the microlens array where individual lens units are randomly distributed across a 20×20 mm² area. We ensure a relatively uniform lens density across the array, with a near unity fill factor and a desired distance range between lenses (mean 1.23 mm, standard deviation 0.09 mm). A negative mold of the microlens array was manufactured through 3D optics printing (Luximprint). We then transferred the pattern into a lens array with 1 mm thick substrate using optical transparent PDMS (SYLGARD® 184, refractive index ~ 1.43) (Fig. S7). We use a blue LED with a 457/50 nm bandpass filter as an illumination source to the object. A fluorescent emission bandpass filter (525/45 nm) is attached to the microlens array, which is mounted on a translation stage to facilitate a fine tuning of the distance between the microlens array and the camera sensor.

**Background suppression through a particle clustering algorithm**

Particle clustering algorithm is developed to suppress the background after the pixel back projection, and is suitable for sparse objects (Fig. S1A and S2). It is based on graph connectivity. We first separate and cluster the 3D volumetric object into isolated groups through a combined operation of thresholding and clustering. The thresholding could remove the ghost objects and the system noise (e.g. stray light, camera readout noise), as they typically have much lower intensity than both the focused light and defocused light from the real object. This essentially cleans up the background between groups of object points isolated from each other. Different groups of object points are then clustered based on their pixel connectivity. For each object group, we reiterate the above process with multiple threshold values so as to remove the defocused light and separate the group into smaller clusters that contain the focused light. For each cluster, we then find the voxel that has peak intensity to represent the position of the object point in focus. Results from different thresholds are unionized to maximize the possibility of separating agminated object points. The clustering method does not require training nor down sampling of large size of image stacks, thus allowing high reconstruction resolution. While this resolution could be arbitrarily small, it is ultimately determined by the meaningful reconstruction voxel size, which depends on system magnification and 3D resolving ability. See more details in Supplementary Section 1.

**Background suppression through a convolutional neural network**

For less sparse 3D objects, we developed a convolutional neural network to suppress the background (Fig. S1A). It slices overlapped objects in depth and picks out focused objects from background light. The network contains five levels of down sampling and up sampling. To train the network, we generate polygon shapes with random intensity and randomly distributed in different depths. We stack ten such slices together to form one set of training data, and in each slice we superimpose the features in current slice and the Gaussian blurred features from its nearby slices (Supplementary Section 2, Fig. S3). The slices have a lower weight when they become farther away. We also add 5% of additive noise on the image. The network is trained with single slice each time with RMSprop optimizer [39] and we use mean squared error as the loss function. The output image from the neural network contains sharper contrast and the defocused features are largely removed.

**ACKNOWLEDGEMENT**
We acknowledge support from National Eye Institute (R21EY029472) and Burroughs Wellcome Fund (Career Award at the Scientific Interface 1015761).

**AUTHOR CONTRIBUTIONS**
Conceptualization, W.Y. and F.T.; Investigation, W.Y., F.T. and J.H.; Data curation: F.T.; Formal analysis: F.T. and W.Y.; Methodology: W.Y. and F.T.; Funding acquisition: W.Y.; Supervision: W.Y.; Writing – Original Draft Preparation: F.T. and W.Y.; Writing – Review & Editing: W.Y., F.T. and J.H.

**COMPLETING INTERESTS**
The authors declare no conflicts of interest.